\documentclass[journal=cmatex,manuscript=letter, layout=twocolumn]{achemso}
\setkeys{acs}{keywords = true}
\usepackage{epsfig}
\usepackage{graphicx}
\usepackage{dcolumn}
\usepackage{color}
\usepackage{natbib}  
\definecolor{blue}{RGB}{0,0,255}
\usepackage{bm}
\usepackage{tabularx}
\usepackage{hyperref}
\usepackage{breakurl}
\hypersetup{colorlinks=true, citecolor=blue, urlcolor=blue, linkcolor=blue}
\newcolumntype{L}[1]{>{\raggedright\arraybackslash}p{#1}}
\newcolumntype{C}[1]{>{\centering\arraybackslash}p{#1}}
\newcolumntype{R}[1]{>{\raggedleft\arraybackslash}p{#1}}
\title{Non-van der Waals honeycomb antiferromagnet SrRu$_2$O$_6$ down to a few layers}

\author{Suvidyakumar Homkar}
\affiliation{Department of Physics, Indian Institute of Science Education and Research, Dr. Homi Bhabha Road, Pune 411008, India.}
\author{Bharat Chand}
\affiliation{Department of Physics, Indian Institute of Science Education and Research, Dr. Homi Bhabha Road, Pune 411008, India.}
\author{Shatruhan Singh Rajput}
\affiliation{Department of Physics, Indian Institute of Science Education and Research, Dr. Homi Bhabha Road, Pune 411008, India.}
\author{Sandeep Gorantla}
\affiliation{{\L}UKASIEWICZ Research Network PORT-Polish Center for Technology Development, ul. Stab{\l}owicka 147, 54-066 Wroc{\l}aw, Poland.}
\author{Tilak Das}
\affiliation{Dipartimento di Scienza dei Materiali, Universit\`a di Milano, Bicocca, via R. Cozzi 55, 20125 Milano, Italy.}
\author{Rohit Babar}
\affiliation{Department of Physics, Indian Institute of Science Education and Research, Dr. Homi Bhabha Road, Pune 411008, India.}
\author{Shivprasad Patil}
\affiliation{Department of Physics, Indian Institute of Science Education and Research, Dr. Homi Bhabha Road, Pune 411008, India.}
\altaffiliation{Centre for Energy Science, Indian Institute of Science Education and Research, Dr. Homi Bhabha Road, Pune 411008, India.}
\author{R\"udiger Klingeler}
\affiliation{Kirchhoff Institute of Physics, Heidelberg University, INF 227, 69120 Heidelberg, Germany.}
\altaffiliation{Centre for Advanced Materials, Heidelberg University, INF 225, 69120 Heidelberg, Germany.}
\author{Sunil Nair}
\affiliation{Department of Physics, Indian Institute of Science Education and Research, Dr. Homi Bhabha Road, Pune 411008, India.}
\altaffiliation{Centre for Energy Science, Indian Institute of Science Education and Research, Dr. Homi Bhabha Road, Pune 411008, India.}
\author{Mukul Kabir}
\affiliation{Department of Physics, Indian Institute of Science Education and Research, Dr. Homi Bhabha Road, Pune 411008, India.}
\altaffiliation{Centre for Energy Science, Indian Institute of Science Education and Research, Dr. Homi Bhabha Road, Pune 411008, India.}
\author{Ashna Bajpai}
\email{ashna@iiserpune.ac.in}
\phone{+91 (20) 2590 8107}
\fax{+91 (20) 2025 1566}
\affiliation{Department of Physics, Indian Institute of Science Education and Research, Dr. Homi Bhabha Road, Pune 411008, India.}
\altaffiliation{Centre for Energy Science, Indian Institute of Science Education and Research, Dr. Homi Bhabha Road, Pune 411008, India.}

\date{\today}

\begin{document}
\begin{abstract}
The current family of experimentally realized two-dimensional magnetic materials consist of 3$d$ transition metals with very weak spin-orbit coupling. In contrast, we report a new platform in a chemically bonded and layered 4$d$ oxide, with strong electron correlations and competing spin-orbit coupling. We synthesize ultra-thin sheets of SrRu$_2$O$_6$ using scalable liquid exfoliation. These exfoliated sheets are characterized by complementary experimental and theoretical techniques. The thickness of the nano-sheets varies between three to five monolayers, and within the first-principles calculations, we show that antiferromagnetism survives in these ultra-thin layers. Experimental data suggest that exfoliation occurs from the planes perpendicular to the $c$-axis as the intervening hexagonal Sr-lattice separates the two-dimensional magnetic honeycomb Ru-layers. The high-resolution transmission electron microscope images indicate that the average inter-atomic spacing between the Ru-layers is slightly reduced, which agrees with the present calculations. The signatures of rotational stacking of the nanosheets are also observed.  Such new two-dimensional platform offers enormous possibilities to explore emergent properties that appear due to the interplay between magnetism, strong correlations and spin-orbit coupling. Moreover, these effects can be further tuned as a function of layer thickness.   
\end{abstract}


\section{Introduction}
                                                      
The continuous rotational symmetry of spins prevents any long-range magnetic ordering in the isotropic Heisenberg model in two-dimension, as described by the celebrated Hohenberg-Mermin-Wagner theorem.~\citep{PhysRev.158.383,PhysRevLett.17.1133} The long-wavelength excitations become gapless, and thus any tendency to magnetic order is destroyed at finite temperature.  However, spins in real materials are far from being isotropic, and the dipolar interaction or relativistic spin-orbit coupling produces magnetic anisotropy, which breaks the continuous rotational invariance of the spins. Therefore, the excitation becomes gapped, and a magnetic phase transition emerges in two-dimension.  Unfortunately, despite its immense importance , it is rare to find two-dimensional magnetic materials with strong spin-orbit coupling.~\citep{Gibertini2019,Gongeaav4450}  

The recent discovery of antiferromagnetic ordering in FePS$_3$ together with the ferromagnetic ordering in CrGeTe$_3$ and CrI$_3$ are the first-ever demonstration of magnetic order in truly two-dimensional materials.~\citep{nanolett.6b03052,Gong,Huang} This breakthrough marks a new era in fundamental research along with enormous application possibilities in two dimensions, and extend functionalities beyond the already existing phases such as ferroelectricity, superconductivity, charge density waves, and various topological states. Apart from these materials, a few other ferromagnetic materials CrCl$_3$, CrBr$_3$, VI$_3$, VSe$_2$, Fe$_3$GeTe$_2$ and MnBi$_2$Te$_2$; and antiferromagnet like NiPS$_3$ have been discovered.~\citep{Kim11131,s41565-019-0565-0,s41567-019-0651-0, adma.201808074,s41565-018-0063-9,s41586-018-0626-9, s41467-018-08284-6} Further, electric manipulation of magnetism has been achieved in these two-dimensional materials, that opens up opportunities in spintronic applications.~\citep{s41565-018-0121-3,s41565-018-0135-x,s41586-018-0626-9} It is important to note here that all the materials that are discovered till date are derived from the van der Waals bulk materials and consists of 3$d$ transition-metals that have weak spin-orbit coupling strength.~\citep{Gibertini2019,Gongeaav4450} Since the anisotropic interactions originating from the spin-orbit coupling opens a gap in the spin-wave spectrum, the corresponding magnetic transition temperature is undesirably low in these materials.~\citep{Lado_2017} Therefore, it would be exciting to discover a new class of two-dimensional magnets with much stronger spin-orbit couplings that couple with strong electron correlation. The interplay between these competing interactions at the comparable energy scales along with the band topology is expected to trigger emergent physics and exciting functionalities in two-dimensional materials.

Transition metal oxides with the 4$d$ and 5$d$ elements are the prime examples of such competing interactions with comparable electron correlation and spin-orbit coupling.~\citep{Khomskii, Cox} Consequently, many exotic physics have been discovered in these oxides in their bulk phases. For example, spin-triplet superconductivity~\citep{JPSJ.81.011009} and emergent Higgs mode oscillations have been discovered in the ruthenium-based materials.~\citep{nphys4077,PhysRevLett.119.067201} 
In the context of two-dimensional magnetism, the difficulty in exfoliation of these non-vdW materials is the main obstacle. While the magnetism in many of these oxides appears in layers, these magnetic layers are chemically bonded with one another, which makes exfoliation a challenging task. 

Another bottleneck that prevents the use of these 4$d$ and 5$d$ transition metal oxides in multifunctional devices is that the magnetic transition temperatures of most of these oxides are far below room temperature. Only a very few 4$d$ and 5$d$ oxides, such as CaTcO$_3$,~\citep{10.1021/ja109431t} SrTcO$_3$,~\citep{PhysRevLett.106.067201} NaOsO$_3$~\citep{PhysRevB.80.161104,PhysRevLett.108.257209} and Sr$_3$OsO$_6$,~\citep{s41467-019-08440-6} have been discovered to have very high magnetic ordering temperatures but none of these materials displays layered magnetic structure. 
Hence, the realization of strongly correlated oxides with high transition temperatures is an area of intense theoretical and experimental investigations.  In this context, the recent discovery of SrRu$_2$O$_6$ with high Neel temperature $T_{\rm N}$ of 560 K is of great importance,~\citep{Nan1,Hiley2} which also holds some exotic electronic properties owing to both localized and itinerant character of electrons.~\citep{Nan3, mazin} Bulk SrRu$_2$O$_6$ is a layered antiferromagnetic insulator with honeycomb magnetic lattice of Ru$^{5+}$ ions and the properties could be further tailored by inducing strain in the lattice to host topological states.~\citep{Nan1,Nan2, Hiley1, Hiley2}  Therefore, the layered SrRu$_2$O$_6$ is a  potential candidate for exfoliation,  and consequently  for experimental realization of two-dimensional magnetism in strongly correlated oxides.  

Here, we report the synthesis of ultra-thin sheets of SrRu$_2$O$_6$ using the scalable technique of liquid exfoliation. These nano-sheets are formed and preserved in the liquid medium of ethanol, which enables easy drop-casting on a suitable substrate for further characterization and device patterning. The ultra-thin sheets are characterized by a wide range of complementary experimental and theoretical techniques. Experimental results indicate stable nano-sheets with the thickness between three to five monolayers. The theoretical calculations suggest the survival of antiferromagnetism in these nano-sheets.  Therefore, these few-layer oxide sheets serve as the new platform beyond the current family of two-dimensional magnetic materials, which additionally host electron correlation and spin-orbit coupling at competing scales.  Moreover, the strongly correlated electrons on the hexagonal lattice with antiferromagnetic interactions provide a fertile playground for emergent phenomena.~\citep{Kane,Haldane,Zhang1}  In this context, SrRu$_2$O$_6$ is also quite similar to the celebrated RuCl$_3$~\citep{Kitaev4,PhysRevB.91.144420} and Na$_2$IrO$_3$ systems,~\citep{Yogesh3} in which Ru or Ir ions form a magnetic hexagon with antiferromagnetic correlations. Both these systems are extensively explored for the possibility of hosting fractionalized Kitaev physics with unconventional excitations.~\citep{Khaliullin, Yogesh1, Yogesh2, Yogesh3,Knolle,Streltsov}

\begin{figure*}[!t]
\includegraphics[width=1\textwidth]{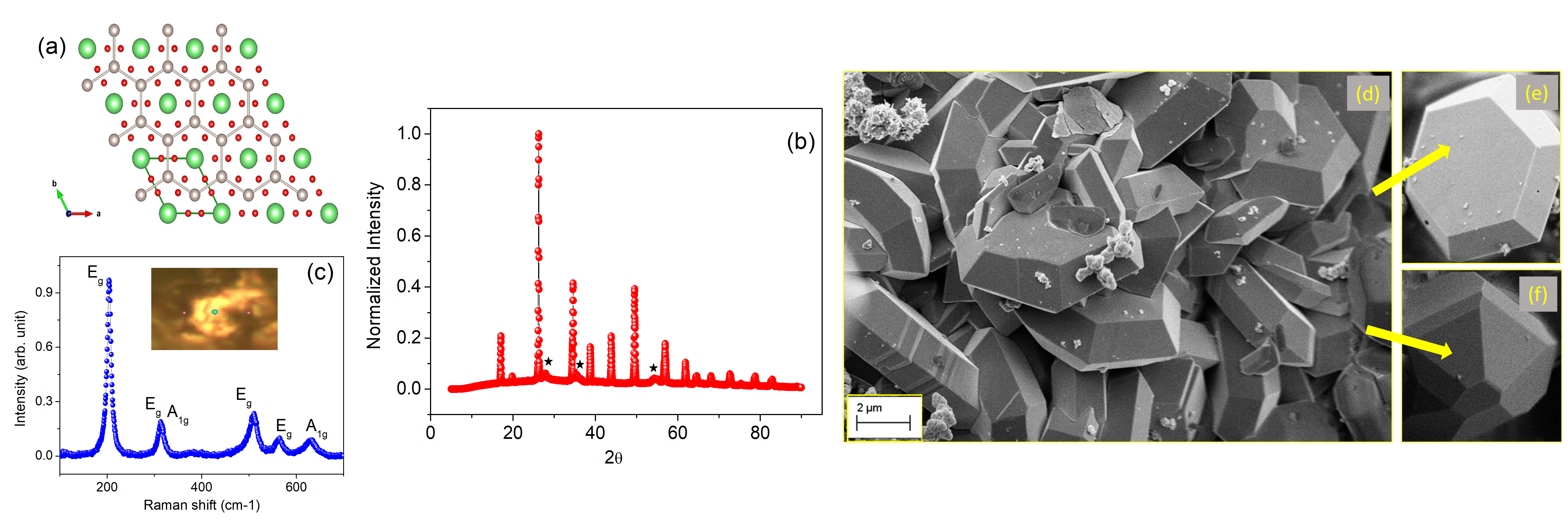}
\caption{(a) Schematic of a honeycomb lattice of Ru ions (gray balls) in the $ab$ plane. The Ru layers are separated by Sr layers (green balls) along the $c$ axis. (b) depicts room temperature powder XRD data for  SrRu$_2$O$_6$ synthesized using the hydrothermal method. Here the intensity has been normalized with the most intense peak of SrRu$_2$O$_6$. A  tiny impurity of RuO$_2$ is marked by stars. (c) shows the Raman data, obtained on a crystallite of SrRu$_2$O$_6$, with the optical image for the crystallite being shown in the inset. The morphology of the sample is depicted in the SEM image of bulk SrRu$_2$O$_6$ as shown in (d), which depicts an overall view of micron-size crystallites forming in regular morphology. (e) and (f) are images of isolated hexagonal crystallites and regular polyhedra.The liquid exfoliation technique has been employed to obtain ultra-thin sheets of SrRu$_2$O$_6$, using these regular shaped  crystallites shown in (d) to (f).}
\label{Figure1}
\end{figure*}

\section{Experimental Techniques}
SrRu$_2$O$_6$ crystallites  are synthesized via the hydrothermal route, using an autoclave from Parr instruments. The morphology and size of as-synthesized bulk  crystallites as well as the nano-sheets of the same, are recorded using a Zeiss Ultra plus FESEM.  Powder X-ray diffraction (XRD) on the crystallites have been conducted using Bruker D8 Advance with Cu $K_\alpha$ radiation ($\lambda$ = 1.54056 \AA). Raman spectroscopy has been done using Horiba instruments equipped with a blue laser ($\lambda$ = 488 nm) and a confocal microscope. The nano-sheets are characterized by Transmission Electron Microscopy  (FEI/TFS Tecnai 20, equipped with LaB6 electron gun, operated at 200 kV). TEM specimens are prepared by drop casting a few $\mu$L  of solution of SrRu$_2$O$_6$ nano-sheets, dispersed and preserved in ethanol, onto Cu TEM grids coated with amorphous carbon support film. The Selected Area Electron Diffraction patterns (SAED) and Fast Fourier Transform (FFT) diffractograms are analyzed by ELDISCA software.~\citep{Gemming} The nano-sheets are further characterized using Atomic Force Microscopy (AFM) using JPK Nanowizard II.  For the synthesis of SrRu$_2$O$_6$ nano-sheets, sonicator (Model Bandelin Sonorex RK 100 H) and Hettich centrifuge (Model MIKRO 200-R) has been used.

The process of liquid exfoliation is routinely employed to form ultra-thin sheets of graphene-like  van der Waals materials, including a wide variety of chalcogenides  such as MoS$_2$.~\citep{Nicolosi1, Griffin, Sean,Backes} This process involves finding a suitable liquid medium in which the bulk sample is sonicated and centrifuged. Here  important synthesis parameters include  (i) sonication  which enables separation of layers which are weakly bonded (ii) centrifugation which enables isolation of thin sheets. The suitable choice of liquid medium  is important, as it prevents the isolated nano-sheets from coagulating again. Time of sonication and centrifugation are also crucial parameters that can be tuned to further refine the thickness distribution as well as size of the nano sheets.~\citep{Nicolosi1, Griffin, Sean,Backes} While a large number of  layered van der Waals materials have been successfully exfoliated using this technique, there are relatively  few reports~\citep{Balan,Sasaki,Sun, Robinson} of exfoliation of magnetic oxides, especially layered ones, with high spin-orbit coupling. Oxides such as SrRu$_2$O$_6$ are quasi two dimensional in the sense of weak inter-layer bonding as compared to in-plane bonding, though both inter-plane and intra-layer bonding are covalent in nature. This anisotropy in bonding strength can thus be exploited to exfoliate ultra-thin sheets of non van der Waals but layered systems,  using the technique of liquid exfoliation.

\section{Results and Discussions}

The compound SrRu$_2$O$_6$, crystallizes in a quasi-2D structure with space group  $P\bar{3}1m$ and lattice parameters $a$ = 0.520573 nm,  $c$ = 0.523454 nm.~\citep{Hiley1} As shown in Figure~\ref{Figure1}(a), the structure of SrRu$_2$O$_6$ comprises of edge-sharing RuO$_6$ layers separated by Sr$^{2+}$ ions. Within the $ab$ plane, the Ru ions are arranged in a honeycomb structure  and Ru moments couple antiferromagnetically both in-plane and out-of-plane. The magnetic moment is 1.30 $\mu_B$/Ru at room temperature and the system possesses the G-type magnetic structure.~\citep{Hiley1, Hiley2} 

\begin{figure}[!t]
\includegraphics[width=0.49\textwidth]{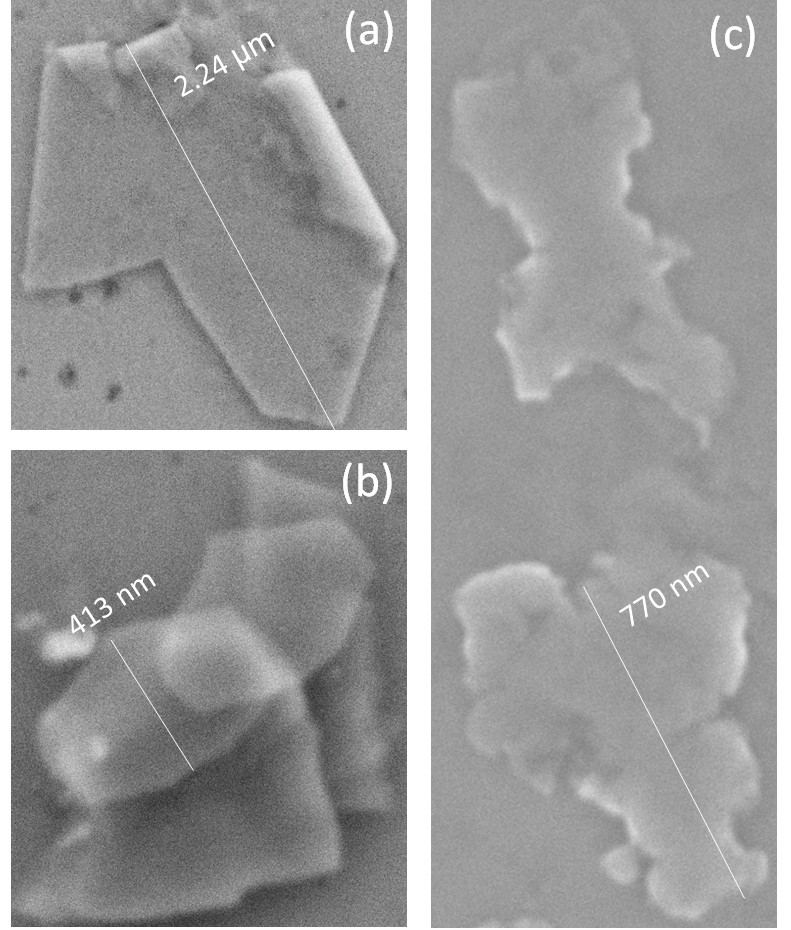}
\caption {(a)--(c) are SEM images of thin flakes of SrRu$_2$O$_6$, post the process of liquid exfoliation, indicating that sheets of various thickness and shapes have been formed. These images are recorded after drop casting the exfoliated sample on Si substrates. }
\label{Figure2}
\end{figure} 

For obtaining the nano-sheets, we first synthesize bulk SrRu$_2$O$_6$ employing the hydrothermal technique described earlier.~\cite{Hiley1} For this purpose, KRuO$_4$ and SrO$_2$ (in stoichiometric ratios) are added in distilled water and stirred, prior to heating the mixture at 200 C for 24 hours in an  autoclave. The precipitate was recovered using a vacuum filtration assembly and washed with diluted HCl, distilled water and acetone. The sample thus obtained was characterized by XRD and Raman, as is shown in Figure~\ref{Figure1}(b) and Figure~\ref{Figure1}(c), respectively.  The  XRD pattern matches well with, JCPDS (04-021-3995) entry assigned for SrRu$_2$O$_6$.~\cite{Hiley1} The lattice parameters derived from XRD data  are  $a$ = 0.5205(7) nm,  $c$ = 0.5219(5) nm. A  tiny amount of  RuO$_2$ phase, JCPDS (65-2824), is also observed as marked by stars in Figure~\ref{Figure1}(b). This impurity  could be systematically reduced by repeated washing the \textit{as-prepared} compound and does not interfere in the exfoliation of SrRu$_2$O$_6$. A characteristic Raman spectra  acquired using blue laser  is shown in the main panel of Figure~\ref{Figure1}(c). The inset shows the optical  image of the crystallite, on which the Raman data has been recorded. The Raman data shown in Figure~\ref{Figure1}(c) are consistent with the previous reports.~\cite{Ponosov} .  Magnetic Characterization on SrRu$_2$O$_6$ crystallites is conducted using SQUID magnetometry ( Quantum Design MPMS) and magnetization as function of temperature from 5K to 600 K  is presented in Supp.Info Figure S1. The magnetization data are in agreement with previous reports~\cite{Hiley1, Hiley2} and further confirm the phase purity of the parent compound used for the process of liquid exfoliation. The morphology of the SrRu$_2$O$_6$ depicting micron-size  crystallites observed  with regular facets, vertices and edges is shown in Figure~\ref{Figure1}(d)-(f). It is to be noted that a large number of crystallites are formed in regular hexagonal shape as is highlighted in Figure~\ref{Figure1}(e).

\begin{figure*}[!t]
\includegraphics[width=1\textwidth]{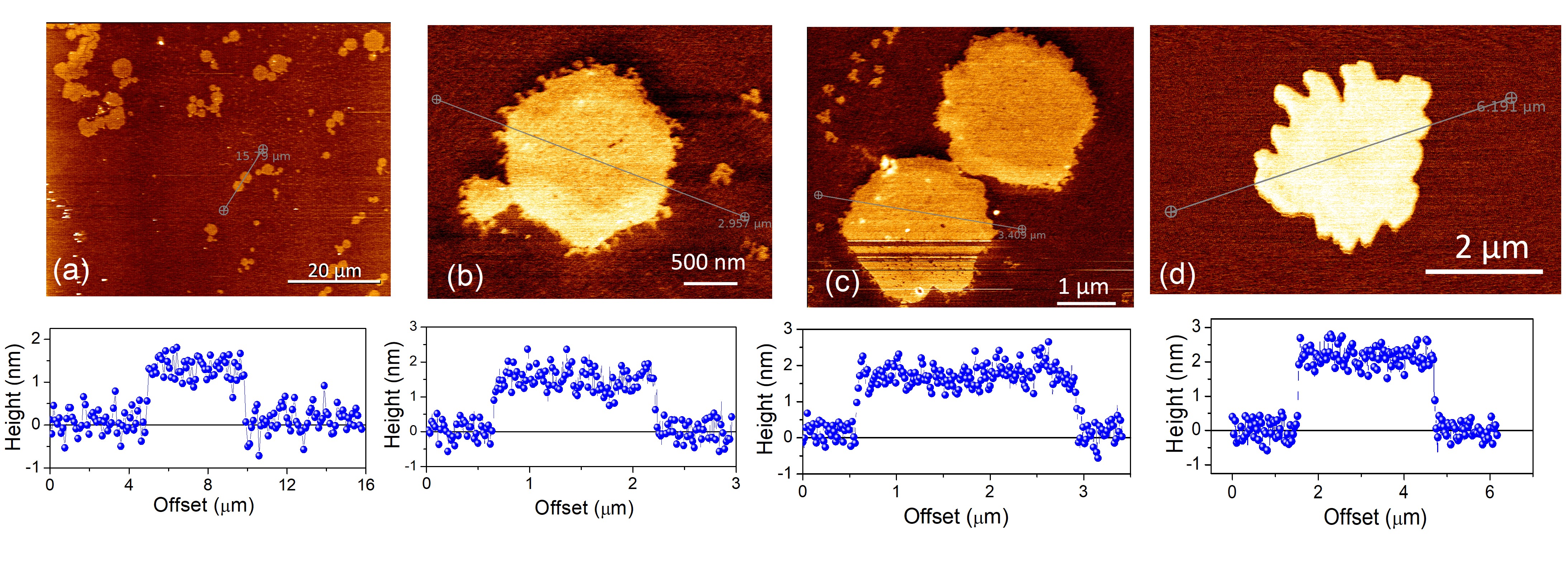}
\caption {(a) - (d) are AFM images of the SrRu$_2$O$_6$ nano sheets of different shape and thickness drop-cast on a mica substrate. (a) cover a broad area, depicting a large number of nano flakes. Both isolated and agglomerated nano-sheets can be clearly seen here. (b) shows a hexagonal shaped nano flake, depicting the morphology of the as-prepared bulk SrRu$_2$O$_6$ shown in Figure~\ref{Figure1}. The  average thickness of the individual  nano-sheet is in the range of 1.5 to 2.2 nm, as is evident from the corresponding line profile shown at the bottom of (a) to (d).} 
\label{Figure3}
\end{figure*}

To obtain nano-sheets of SrRu$_2$O$_6$, the first task is to determine a suitable  liquid medium, in which the bulk crystallites can be dispersed. This solution , containing the bulk crystallites (in an optimized liquid-medium and crystallite ratio) is then subjected to sonication for  mechanically isolating the weakly bonded layers. After trying a number of solvents, we found that ethanol is most suitable  for obtaining ultra-thin sheets of SrRu$_2$O$_6$. For this purpose, as-prepared  SrRu$_2$O$_6$  crystallites  were dispersed in ethanol. Usage of ethanol as a solvant for the process of liquid exfoliation has certain advantages. For instance, nano-sheets preserved in ethanol are easy to drop-cast on a suitable substrate. As ethanol evaporates, this provides an opportunity for easy device patterning.  In order to synthesize ultra-thin sheets, we first make a solution of SrRu$_2$O$_6$ crystallites in ethanol, typically in a 1:10 ratio. This solution is subjected to 4 hours of bath and probe sonication, following which the solution is kept undisturbed for 24 hours. The top portion of the solution is collected and centrifuged at 2000 rpm for 30 minutes. These experimental parameters ( such and time of sonication or centrifugation etc.) can be further refined to obtain thinner nano-sheets. For instance, multiple rounds of sonication and centrifuge usually leads to the collection of relatively thinner sheets in the top portion of the centrifuge vial. The centrifuge vial is also kept undisturbed for a few hours  prior to retrieving a few $\mu$L from the top layer using a micro pippet. This refined solution is then drop-cast onto the various suitable substrates for further characterization.  

The Figures~\ref{Figure2}(a)-(c) displays SEM images of the nano-sheets thus formed. The images are recorded by drop-casting a few $\mu$L of the solution, that contains dispersed nano sheets,  on a Si substrate. As evident from Figure~\ref{Figure2}, the sheets are formed in various shapes and size, for which a few representatives are shown here. Most importantly, we observe a significant fraction of sheets  retain the shape of regular polyhedra of \textit{as-formed} micro-crystals shown in Figure~\ref{Figure1}(d). Both isolated as well as a bunch of nano-sheets (which lie on top of each other) can be found. The lateral length of such nano-sheets varied from few tens of nano meters to a few 100 nm and again appears to correlate with dimensions of the bulk SrRu$_2$O$_6$ crystallites, from which these sheets have been exfoliated. 

While the presence of sheets with different thicknesses can be gauged from the contrast in the SEM images, we performed AFM measurements for determining the average thickness of the nano sheets. For this purpose,  solution containing dispersed nano -sheets is drop-casted on a freshly cleaved mica substrate. The Figures~\ref{Figure3}(a)-(d) show some representative AFM images on the nano-sheets of SrRu$_2$O$_6$. The Figure~\ref{Figure3}(a) shows a broad area AFM image, displaying a large number of sheets  of various shape and size on the substrate - also demonstrating the scalability which the technique of liquid exfoliation provides. This also grants us  means to pick the sheets with desired thickness for future magnetotransport measurements. Many of the individual nano-sheets, such as shown in Figure~\ref{Figure3}(a), retain the shape of the bulk crystallites, used in the exfoliation.  This is evident from  AFM images shown in Figure~\ref{Figure3}(b)-(c), which show some neatly formed hexagonal sheets, and a rather distorted hexagon.  A fairly arbitrary shape sheet is shown in Figure~\ref{Figure3}(d) which bears striking  similarities in SEM images shown for arbitrary shaped sheets in Figure~\ref{Figure3}(c).

\begin{figure*}[!t]
\includegraphics[width=0.95\textwidth]{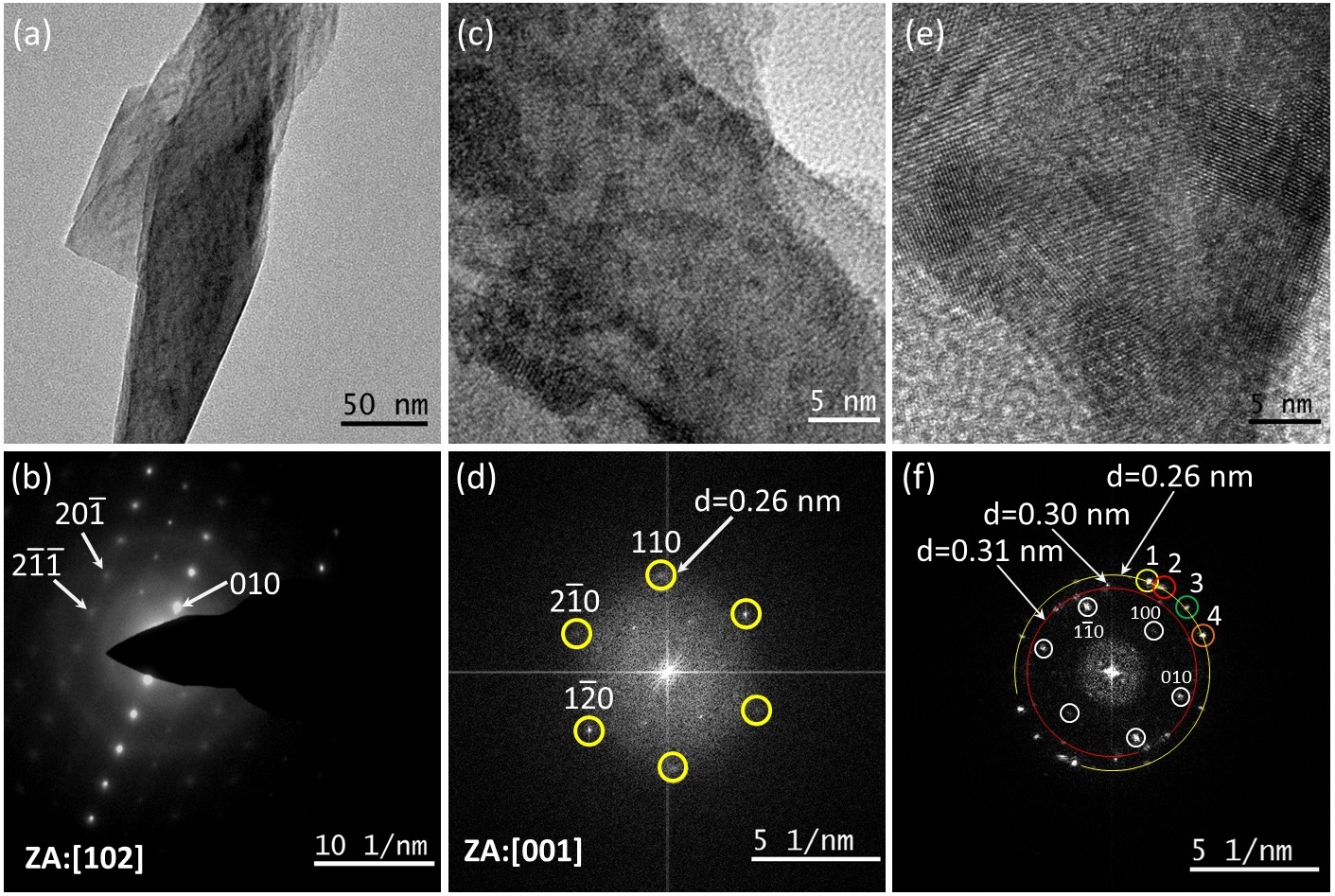}
\caption{TEM, SAED and HRTEM characterization of the exfoliated SrRu${_2}$O${_6}$ flakes. (a) and (b) show an overview TEM image and corresponding SAED pattern of an exfoliated flake respectively. The diffraction pattern is indexed in agreement with [102] zone axis of SrRu${_2}$O${_6}$. (c) and (d) show a HRTEM image and corresponding  FFT of an area at the edge of a few-layer flake respectively. The hexagonal symmetry of the spots in the FFT, marked by circles, is indexed in agreement with [001] zone axis of SrRu${_2}$O${_6}$. (e) and (f)  show HRTEM image and corresponding FFT respectively of an area at the edge of another few-layer flake. In this FFT, multiple spots with 0.26 nm and 0.31 nm d-spacing are marked with outer and inner partial circle respectively. The spot corresponding to 0.30 nm d-spacing (bulk Ru-Ru spacing in [001] projection) is marked by an arrow.  The spots in FFT marked by numbered circles originate from four different regions at the edges of the flake in the corresponding HRTEM image, where (110) planes are rotated with respect to each other. The inner skewed hexagonal spots close to center of FFT, marked by white circle markers are indexed with [001] zone axis SAED of SrRu${_2}$O${_6}$ crystal structure. The (1-10) and (010) spots in the FFT are observed to deviate away from bulk crystal SAED pattern positions. This may be due to shrinking of the d-spacing of these planes in the real space after exfoliation.} 
\label{Figure4}
\end{figure*}
	
From the AFM images presented in Figure~\ref{Figure3}(a), we estimated average thickness of sheets $\sim$ 1.31 nm ($\pm$ 0.12 nm), taking into account about 30  different sheets.  The  thickness of the hexagonal shaped sheets shown in Figure~\ref{Figure3}(b)-(c) is $\sim$ 1.5 nm, whereas it is $\sim$ 2.2 nm for the sheet shown in Figure 3d. Considering that the process of liquid exfoliation enables termination of sheets from Sr layer, the number of layers in perfectly stacked sheets can be gauged by looking into the lattice parameter $c$ = 0.523 nm, corresponding to bulk SrRu$_2$O$_6$.  This approach  suggests  three to five monolayers, considering a significant reduction in lattice parameter $c$ from  its bulk value. This is especially interesting, as  prior density functional theory (DFT) calculations have suggested that strain mediated topological properties could be observed in  samples with certain thickness.~\citep{Nan2}
		
We also observed that repeating the entire process of sonication and centrifugation (after leaving the dispersed nano-sheets for months in the vial) resulted in narrower distribution of thickness. The dispersed nano sheets in the solution, even after prolonged time-periods, retained the morphology, such as shown in  Figure~\ref{Figure2}. Furthermore,  repeated process of sonication and centriguation performed using  the solution preserved in the same vial led to the observation of relatively thinner sheets. In all these samples, many regions with partially stacked sheets, such as shown in Figure~\ref{Figure2}(b), could also be found.  Thus it is evident that the each step involved in the process of liquid exfoliation, including the time of sonication and centrifugation  can be further refined to obtain either monolayers or fully/partially stacked  stacked nano-sheets. 
	
Exploring novel electronic states in van der Waals driven 2D materials, especially graphene  is an area of extensive research~\citep{Kim3,Kim4,Kim5}.  However, antiferromagnetic interactions in a system with magnetic honeycomb and large spin orbit coupling  is a testbed for exploring topological states. This is especially so in the case of SrRu${_2}$O${_6}$, where DFT predicts strain mediated topological state and band inversion when the crystallographic $c$ axis could be significantly tailored, leaving the $a$ and $b$ lattice parameters relatively unchanged \cite{Nan2}. It is evident here that the nano-sheets of SrRu$_2$O$_6$ produced using the technique of liquid exfoliation possibly provides a means to obtain ultra-thin 2D sheets, of various thickness and lateral lengths. From SEM and AFM images shown in Figure~\ref{Figure2} and Figure~\ref{Figure3}, it is evident that these 2D sheets can range from a few 100 nm to a few microns. Thus, depending on the size and the thickness of the nano-sheet, both $a$ as well as $c$ lattice parameters can be strained anisotropically. For instance, a monolayer  which extends to a few microns in length, the $c$  parameter is likely to be more strained than  $a$  lattice parameter.  Experimental realization  of such nano-sheets with varying thickness and lateral length  also enables a possibility in which a particular lattice parameter (perpendicular to the weakly bonded layers) can be tuned, without significantly disturbing the  (in-plane) lattice parameters.  Such  2D sheets of SrRu${_2}$O${_6}$ can provide an opportunity to  test the theoretical predictions, such as discussed in reference 32.

A representative overview TEM image of an individual flake is shown in Figure~\ref{Figure4}(a). It shows portion of a typical folded sheet of SrRu$_2$O$_6$, which is about 100 nm wide and a few 100 nm long. The corresponding SAED pattern obtained from this sheet, as shown in Figure~\ref{Figure4}(b), could be indexed in agreement with the [102] zone axis diffraction pattern of SrRu$_2$O$_6$. Figure~\ref{Figure4}(c) shows HRTEM image from an edge of the exfoliated nano-flake. In this image, atleast three different nano sheets edges can be seen. The FFT of this image in Figure~\ref{Figure4}(d) shows the spots in agreement with the
hexagonal spot pattern of [001] zone axis Selected Area Electron diffraction (SAED) pattern of SrRu$_2$O$_6$. The measured d-spacing from these spots, 0.26 nm, is in good agreement with the (110), (2-10) and (1-20) planes spacing of d= 0.2603 nm (from JCPDS card entry  04-021-3995 for SrRu$_2$O$_6$). This  observation  not only confirms that these crystalline  nano -sheets have retained the bulk phase but also suggests that the nano-sheets may have been exfoliated from planes perpendicular to the $c$-axis of this system.

Figure~\ref{Figure4}(e) shows HRTEM image from a region of another flake, where some regions with moire pattern like contrast are observed. The corresponding FFT, Figure~\ref{Figure4}(f), shows FFT multiple spots with 0.26 nm d-spacing (supplementary figure TEM-S2), i.e. (110) planes as marked with outer partial circle. The spots that lie on this ring, as marked by numbered circles, originate from four different regions at the edges of the flake in the corresponding HRTEM image where (110) planes are rotated with respect to each other. Presence of multiple spots along a ring with constant d-spacings is typical of SAED/FFT pattern from a polycrystalline sample with randomly oriented grains. However, this does not appear to be the case here and is reminiscent of prior observations in few layer 2D materials like graphene, where similar FFT patterns were observed due to rotational stacking faults.~\citep{Warner} To satisfy this condition each spot in the marked numbered circle should clearly show a set of corresponding hexagonal spots which are not clearly observed in Figure~\ref{Figure4}(f). Thus, we can speculate that the  pattern observed here may point towards the presence of stacked exfoliated sheets with different azimuthal rotation between them. However, it is worth noting that such a possibility of rotationally stacked nano-sheets is highly likely in this sample, as this tendency is also observed seen in SEM images, Figure~\ref{Figure2}(b). 

In Figure~\ref{Figure4}(f), the spot corresponding to 0.30 nm d-spacing (bulk Ru-Ru spacing in [001] projection of SrRu$_2$O$_6$) is marked by an arrow, below which  the multiple spots with constant d-spacing are marked with inner partial circles. The average spacing of the spots on this ring is measured to be 0.31 nm (supplementary figure TEM-S3). This observation indicate slight in plane shrinking of Ru-Ru interatomic spacing in the exfoliated nanosheets. Finally, in Figure~\ref{Figure4}(f), the six spots close to the center of FFT (as marked by white circle markers) show a distorted hexagonal arrangement. They are indexed with [001] zone axis SAED of SrRu$_2$O$_6$ crystal structure. The (1-10) and (010) spots in the FFT are observed to deviate away from their expected positions in simulated [001] zone axis SAED pattern positions. The measured d-spacing of these planes show an average spacing of 0.365 nm as compared to the bulk value of 0.45 nm. This indicates shrinking of the d-spacing of these planes after exfoliation. Also the measured d-spacing of 100 planes in the FFT (0.43 nm) indicate slight decrease of the inter planar spacing between them (supplementary Figure S4). We believe that strain effects in lattice parameters upon exfoliating a non van der Waals but layered system such as SrRu$_2$O$_6$, are likely to be larger. TEM data presented here is indicative of large strain effects as observed in these nano sheets. We also recorded EDX in TEM on a number of  2D nano-sheets, a representative of which is shown as Figure S5 in the supplementary information. 

\begin{figure}[!t] 
\includegraphics[width=0.49\textwidth]{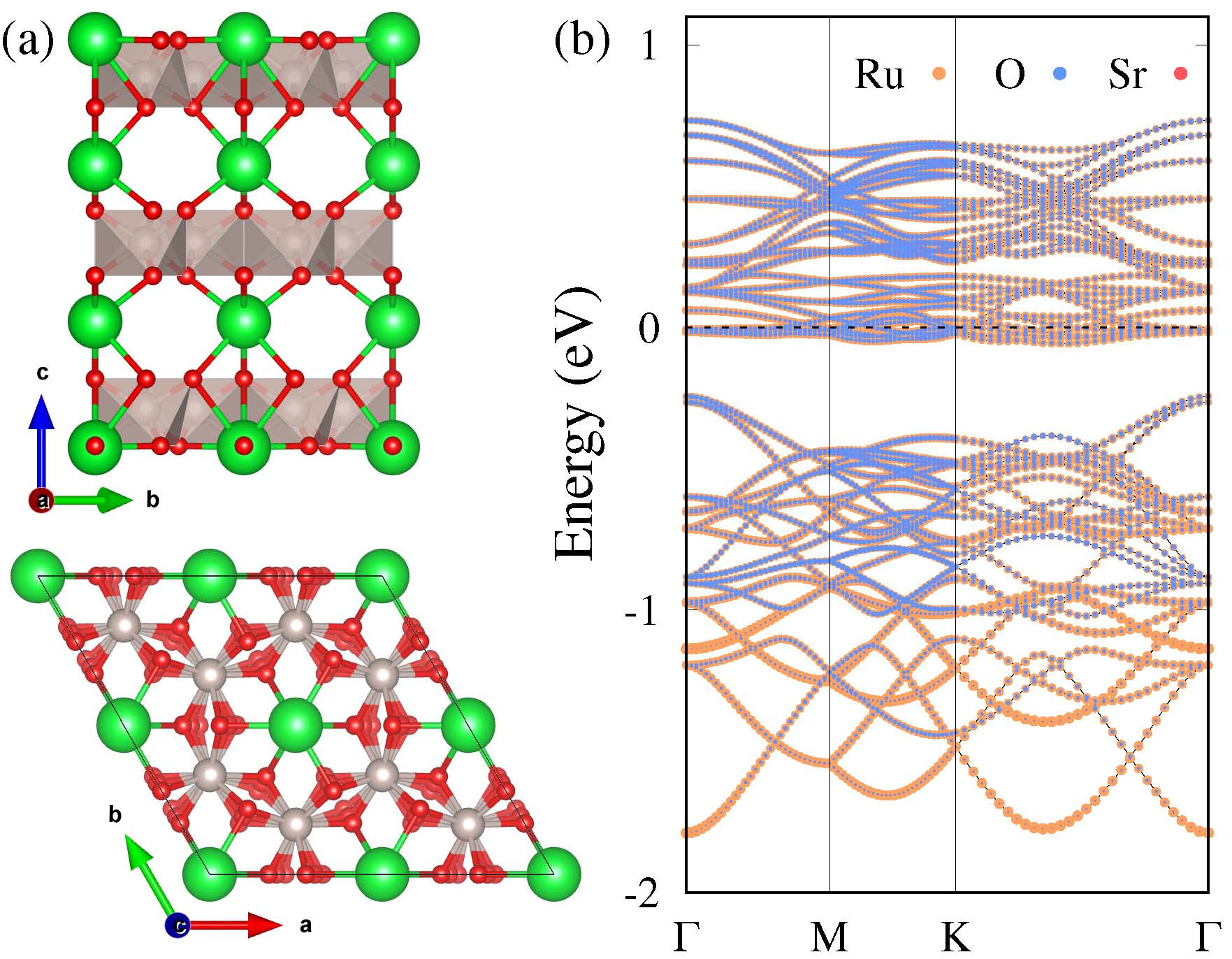}
\caption {(a) Top and side views of the optimized lattice of 3L SrRu$_2$O$_6$ ultrathin film. The surface Sr-layer relaxes toward the RuO$_6$ octahedra, and the O-positions indicate distortion in the RuO$_6$ octahedra at the surface. (b) Band dispersion of the 3L SrRu$_2$O$_6$ in the G-type antiferromagnetic magnetic structure exhibit metallic character. The Fermi energy is set to $E=0$. The metallic solution remains invariant in the non-magnetic state, and also when the spin-orbit coupling is incorporated in the calculation. }
\label{Figure5}
\end{figure}

To gain further insights,  about the electronic structure, we characterize the SrRu$_2$O$_6$ nano-sheets within the first-principles density functional theory (DFT) calculations as implemented in the Vienna Ab Initio Simulation Package.~\cite{Kresse1, Kresse2} The wave function is described within the projector augmented wave formalism with 500 eV kinetic energy cut-off.~\cite{Blochl} The exchange-correlation is treated with the Perdew-Burke-Ernzerhof functional,~\cite{Perdew} and the spin-orbit coupling is treated self-consistently. A large vacuum of 17 \AA\ is maintained perpendicular to the surface of the few-layer SrRu$_2$O$_6$ nano-sheet to minimize the spurious interaction between the periodic images. The structure is completely relaxed until the forces reduce below $\pm$0.01 eV/\AA\ threshold. The Brillouin zone is sampled using a 9$\times$9$\times$9 and 9$\times$9$\times$1  $k$-mesh for the bulk and few-layered samples, respectively.
			
Before we discuss the structural, electronic and magnetic properties of the ultrathin SrRu$_2$O$_6$ nano-sheet, we revisit the intriguing bulk phase within the same theoretical hierarchy. The calculated lattice parameters ($a$ = 5.29 and $c$ = 5.26 \AA) are in good agreement with the experimental values calculated from the neutron diffraction data~\cite{Hiley1}. The Ru$^{5+}$-ions in the hexagonal $P\bar{3}1m$ lattice of  SrRu$_2$O$_6$ are octahedrally coordinated with the oxygen ions, which supports the half-filled $t_{2g}$-manifold with 3 $\mu_B$ expected moment. However, due to the strong hybridization between Ru-$t_{2g}$ and O-2$p$ orbitals, the Ru-moment is substantially reduced to 1.38 $\mu_B$, in agreement with prior experimental and theoretical results.~\cite{Tian, Hiley2, Suzuki1, David, Streltsov, Hariki, Nan3} The two-dimensional Ru-spins exhibit G-type antiferromagnetic with semiconducting electronic structure. The corresponding non-magnetic solution indicates a small bandgap of 68 meV in the present calculation, while the gap increases with the Ru-moment~\cite{Streltsov}. We find the G-type antiferromagnetic SrRu$_2$O$_6$ to have a higher 0.44 eV gap, and the inclusion of spin-orbit coupling does not further influence the gap (0.41 eV). 
		
The structure of the Sr-terminated 3L SrRu$_2$O$_6$ sheet is fully relaxed [Figure~\ref{Figure5}(a)],  and the in-plane lattice parameter $a$ is slightly stretched (5.38 \AA), in comparison to the corresponding bulk value. In contrast, the perpendicular lattice parameter $c$ is critically contracted in 3L SrRu$_2$O$_6$ (4.27 \AA~ compared to 5.26 \AA ~in bulk) due to the relaxation of surface Sr-layer toward the RuO-octahedra, which is in remarkable agreement with the experimental observations, particularly  the AFM  and TEM measurements presented in Figure~\ref{Figure3}, Figure~\ref{Figure4} and supplementary Figure S3. These data bring out significant strain effects in the interlayer spacing,  which leads to associated compression or expansion in the lattice parameters of SrRu$_2$O$_6$ nano-sheets. 
		
Similar to the bulk, the ultra-thin SrRu$_2$O$_6$ sheet exhibits G-type antiferromagnetic ordering, which is 30 meV/Ru lower in energy compared to the corresponding ferromagnetic state. The individual Ru-moments decreases to 1.32 $\mu_B$ at the middle RuO-layer, which further reduces at the surface due to RuO$_6$ distortion. Surprisingly, an insulator to metal transition is observed while the thickness is reduced to few-layer [Figure~\ref{Figure5}(b)]. The 3L SrRu$_2$O$_6$ sheet is found to be metallic, which is mostly contributed by the $p-d$ hybridization at the distorted surface RuO$_6$ octahedra. The predicted (antiferro) magnetism in the ultrathin SrRu$_2$O$_6$ nano-sheets will have implications in the search for magnetism in the two-dimension, that is typically observed in the van der Waals systems.~\citep{nanolett.6b03052,Gong,Huang,Gibertini2019,Gongeaav4450}

\section{Conclusions}
In conclusion, we demonstrate a new platform to realize two-dimensional magnetism in non-van der Waals and magnetically layered 4$d$ oxide with strong electron correlation and competing spin-orbit coupling. We exfoliate the ultra-thin nano-sheets of SrRu$_2$O$_6$ using the scalable technique of liquid exfoliation.  While the method of liquid exfoliation was primarily meant for van der Waals materials like graphene, we show that the same technique can be successfully employed to obtain nano-sheets of inherently layered magnetic oxides such as SrRu$_2$O$_6$, with anisotropic in-plane and out-of-plane bond strengths. These nano-sheets could be obtained and preserved using ethanol as a medium. This enables one to easily drop-cast them on a suitable substrate for characterization as well as on patterned substrates for magneto-transport measurements. Scanning and Transmission electron microscopy, as well as Atomic force microscopy, have been used to characterize these nano-sheets. Within the complementary first-principles calculations, we show that antiferromagnetism survives in these ultra-thin nano-sheets. Experimental realization of  two-dimensional sheets of SrRu$_2$O$_6$  which has a graphene-like  magnetic honeycomb,  along with competing electronic correlations and spin-orbit coupling offers tremendous potential to investigate emergent phenomena in this particular type of two-dimensional magnets. The present study will incite further attention in this regard.     
			  
\section{Acknowledgments} 
Authors acknowledge Prof. A.K.Nigam for useful discussions,  Mr. Rudheer Bapat and Mr. J. Parmar  for TEM measurements and Mr. A. Shetty for SEM measurements. A.B. acknowledges Department of Science and Technology (DST), India for funding support through a Ramanujan Grant. A.B., S.N., M.K., and S.P. acknowledge DST Nanomission Thematic Unit Program for financial support. R.K., S.N. and A.B. acknowledge funding support through Joint DST-DFG projects INT/FRG/DFG/P-06/2017 and KL1824/11-1. M.K. acknowledges funding from the Science and Engineering Research Board through EMR/2016/006458 grant.

\bibliography{Bibliography}

\providecommand{\latin}[1]{#1}
\makeatletter
\providecommand{\doi}
  {\begingroup\let\do\@makeother\dospecials
  \catcode`\{=1 \catcode`\}=2 \doi@aux}
\providecommand{\doi@aux}[1]{\endgroup\texttt{#1}}
\makeatother
\providecommand*\mcitethebibliography{\thebibliography}
\csname @ifundefined\endcsname{endmcitethebibliography}
  {\let\endmcitethebibliography\endthebibliography}{}
\begin{mcitethebibliography}{67}
\providecommand*\natexlab[1]{#1}
\providecommand*\mciteSetBstSublistMode[1]{}
\providecommand*\mciteSetBstMaxWidthForm[2]{}
\providecommand*\mciteBstWouldAddEndPuncttrue
  {\def\EndOfBibitem{\unskip.}}
\providecommand*\mciteBstWouldAddEndPunctfalse
  {\let\EndOfBibitem\relax}
\providecommand*\mciteSetBstMidEndSepPunct[3]{}
\providecommand*\mciteSetBstSublistLabelBeginEnd[3]{}
\providecommand*\EndOfBibitem{}
\mciteSetBstSublistMode{f}
\mciteSetBstMaxWidthForm{subitem}{(\alph{mcitesubitemcount})}
\mciteSetBstSublistLabelBeginEnd
  {\mcitemaxwidthsubitemform\space}
  {\relax}
  {\relax}

\bibitem[Hohenberg(1967)]{PhysRev.158.383}
Hohenberg,~P.~C. Existence of Long-Range Order in One and Two Dimensions.
  \emph{Phys. Rev.} \textbf{1967}, \emph{158}, 383--386\relax
\mciteBstWouldAddEndPuncttrue
\mciteSetBstMidEndSepPunct{\mcitedefaultmidpunct}
{\mcitedefaultendpunct}{\mcitedefaultseppunct}\relax
\EndOfBibitem
\bibitem[Mermin and Wagner(1966)Mermin, and Wagner]{PhysRevLett.17.1133}
Mermin,~N.~D.; Wagner,~H. Absence of Ferromagnetism or Antiferromagnetism in
  One- or Two-Dimensional Isotropic Heisenberg Models. \emph{Phys. Rev. Lett.}
  \textbf{1966}, \emph{17}, 1133--1136\relax
\mciteBstWouldAddEndPuncttrue
\mciteSetBstMidEndSepPunct{\mcitedefaultmidpunct}
{\mcitedefaultendpunct}{\mcitedefaultseppunct}\relax
\EndOfBibitem
\bibitem[Gibertini \latin{et~al.}(2019)Gibertini, Koperski, Morpurgo, and
  Novoselov]{Gibertini2019}
Gibertini,~M.; Koperski,~M.; Morpurgo,~A.~F.; Novoselov,~K.~S. Magnetic 2D
  materials and heterostructures. \emph{Nat. Nanotechnol.} \textbf{2019},
  \emph{14}, 408--419\relax
\mciteBstWouldAddEndPuncttrue
\mciteSetBstMidEndSepPunct{\mcitedefaultmidpunct}
{\mcitedefaultendpunct}{\mcitedefaultseppunct}\relax
\EndOfBibitem
\bibitem[Gong and Zhang(2019)Gong, and Zhang]{Gongeaav4450}
Gong,~C.; Zhang,~X. Two-dimensional magnetic crystals and emergent
  heterostructure devices. \emph{Science} \textbf{2019}, \emph{363}\relax
\mciteBstWouldAddEndPuncttrue
\mciteSetBstMidEndSepPunct{\mcitedefaultmidpunct}
{\mcitedefaultendpunct}{\mcitedefaultseppunct}\relax
\EndOfBibitem
\bibitem[Lee \latin{et~al.}(2016)Lee, Lee, Ryoo, Kang, Kim, Kim, Park, Park,
  and Cheong]{nanolett.6b03052}
Lee,~J.-U.; Lee,~S.; Ryoo,~J.~H.; Kang,~S.; Kim,~T.~Y.; Kim,~P.; Park,~C.-H.;
  Park,~J.-G.; Cheong,~H. Ising-Type Magnetic Ordering in Atomically Thin
  FePS3. \emph{Nano Lett.} \textbf{2016}, \emph{16}, 7433--7438\relax
\mciteBstWouldAddEndPuncttrue
\mciteSetBstMidEndSepPunct{\mcitedefaultmidpunct}
{\mcitedefaultendpunct}{\mcitedefaultseppunct}\relax
\EndOfBibitem
\bibitem[Gong \latin{et~al.}(2017)Gong, Li, Li, Ji, Stern, Xia, Cao, Bao, Wang,
  Wang, Qiu, Cava, Louie, Xia, and Zhang]{Gong}
Gong,~C.; Li,~L.; Li,~Z.; Ji,~H.; Stern,~A.; Xia,~Y.; Cao,~T.; Bao,~W.;
  Wang,~C.~e.; Wang,~Y.; Qiu,~Z.~Q.; Cava,~R.~J.; Louie,~S.~G.; Xia,~J.;
  Zhang,~X. Discovery of intrinsic ferromagnetism in two-dimensional van der
  Waals crystals. \emph{Nature} \textbf{2017}, \emph{546}, 265\relax
\mciteBstWouldAddEndPuncttrue
\mciteSetBstMidEndSepPunct{\mcitedefaultmidpunct}
{\mcitedefaultendpunct}{\mcitedefaultseppunct}\relax
\EndOfBibitem
\bibitem[Huang \latin{et~al.}(2017)Huang, Clark, Navarro-Moratalla, Klein,
  Cheng, Seyler, Zhong, Schmidgall, McGuire, Cobden, Yao, Xiao,
  Jarillo-Herrero, and Xu]{Huang}
Huang,~B.; Clark,~G.; Navarro-Moratalla,~E.; Klein,~D.~R.; Cheng,~R.;
  Seyler,~K.~L.; Zhong,~D.; Schmidgall,~E.; McGuire,~M.~A.; Cobden,~D.~H.;
  Yao,~W.; Xiao,~D.; Jarillo-Herrero,~P.; Xu,~X. Layer-dependent ferromagnetism
  in a van der Waals crystal down to the monolayer limit. \emph{Nature}
  \textbf{2017}, \emph{546}, 270\relax
\mciteBstWouldAddEndPuncttrue
\mciteSetBstMidEndSepPunct{\mcitedefaultmidpunct}
{\mcitedefaultendpunct}{\mcitedefaultseppunct}\relax
\EndOfBibitem
\bibitem[Kim \latin{et~al.}(2019)Kim, Yang, Li, Jiang, Jin, Tao, Nichols,
  Sfigakis, Zhong, Li, Tian, Cory, Miao, Shan, Mak, Lei, Sun, Zhao, and
  Tsen]{Kim11131}
Kim,~H.~H. \latin{et~al.}  Evolution of interlayer and intralayer magnetism in
  three atomically thin chromium trihalides. \emph{Proc. Natl. Acad. Sci. USA}
  \textbf{2019}, \emph{116}, 11131--11136\relax
\mciteBstWouldAddEndPuncttrue
\mciteSetBstMidEndSepPunct{\mcitedefaultmidpunct}
{\mcitedefaultendpunct}{\mcitedefaultseppunct}\relax
\EndOfBibitem
\bibitem[Wang \latin{et~al.}(2019)Wang, Gibertini, Dumcenco, Taniguchi,
  Watanabe, Giannini, and Morpurgo]{s41565-019-0565-0}
Wang,~Z.; Gibertini,~M.; Dumcenco,~D.; Taniguchi,~T.; Watanabe,~K.;
  Giannini,~E.; Morpurgo,~A.~F. Determining the phase diagram of atomically
  thin layered antiferromagnet CrCl3. \emph{Nat. Nanotechnol.} \textbf{2019},
  \emph{14}, 1116--1122\relax
\mciteBstWouldAddEndPuncttrue
\mciteSetBstMidEndSepPunct{\mcitedefaultmidpunct}
{\mcitedefaultendpunct}{\mcitedefaultseppunct}\relax
\EndOfBibitem
\bibitem[Klein \latin{et~al.}(2019)Klein, MacNeill, Song, Larson, Fang, Xu,
  Ribeiro, Canfield, Kaxiras, Comin, and Jarillo-Herrero]{s41567-019-0651-0}
Klein,~D.~R.; MacNeill,~D.; Song,~Q.; Larson,~D.~T.; Fang,~S.; Xu,~M.;
  Ribeiro,~R.~A.; Canfield,~P.~C.; Kaxiras,~E.; Comin,~R.; Jarillo-Herrero,~P.
  Enhancement of interlayer exchange in an ultrathin two-dimensional magnet.
  \emph{Nat. Phys.} \textbf{2019}, \emph{15}, 1255--1260\relax
\mciteBstWouldAddEndPuncttrue
\mciteSetBstMidEndSepPunct{\mcitedefaultmidpunct}
{\mcitedefaultendpunct}{\mcitedefaultseppunct}\relax
\EndOfBibitem
\bibitem[Kong \latin{et~al.}(2019)Kong, Stolze, Timmons, Tao, Ni, Guo, Yang,
  Prozorov, and Cava]{adma.201808074}
Kong,~T.; Stolze,~K.; Timmons,~E.~I.; Tao,~J.; Ni,~D.; Guo,~S.; Yang,~Z.;
  Prozorov,~R.; Cava,~R.~J. VI3?a New Layered Ferromagnetic Semiconductor.
  \emph{Adv. Mater.} \textbf{2019}, \emph{31}, 1808074\relax
\mciteBstWouldAddEndPuncttrue
\mciteSetBstMidEndSepPunct{\mcitedefaultmidpunct}
{\mcitedefaultendpunct}{\mcitedefaultseppunct}\relax
\EndOfBibitem
\bibitem[Bonilla \latin{et~al.}(2018)Bonilla, Kolekar, Ma, Diaz, Kalappattil,
  Das, Eggers, Gutierrez, Phan, and Batzill]{s41565-018-0063-9}
Bonilla,~M.; Kolekar,~S.; Ma,~Y.; Diaz,~H.~C.; Kalappattil,~V.; Das,~R.;
  Eggers,~T.; Gutierrez,~H.~R.; Phan,~M.-H.; Batzill,~M. Strong
  room-temperature ferromagnetism in VSe2 monolayers on van der Waals
  substrates. \emph{Nat. Nanotechnol.} \textbf{2018}, \emph{13}, 289--293\relax
\mciteBstWouldAddEndPuncttrue
\mciteSetBstMidEndSepPunct{\mcitedefaultmidpunct}
{\mcitedefaultendpunct}{\mcitedefaultseppunct}\relax
\EndOfBibitem
\bibitem[Deng \latin{et~al.}(2018)Deng, Yu, Song, Zhang, Wang, Sun, Yi, Wu, Wu,
  Zhu, Wang, Chen, and Zhang]{s41586-018-0626-9}
Deng,~Y.; Yu,~Y.; Song,~Y.; Zhang,~J.; Wang,~N.~Z.; Sun,~Z.; Yi,~Y.; Wu,~Y.~Z.;
  Wu,~S.; Zhu,~J.; Wang,~J.; Chen,~X.~H.; Zhang,~Y. Gate-tunable
  room-temperature ferromagnetism in two-dimensional Fe3GeTe2. \emph{Nature}
  \textbf{2018}, \emph{563}, 94--99\relax
\mciteBstWouldAddEndPuncttrue
\mciteSetBstMidEndSepPunct{\mcitedefaultmidpunct}
{\mcitedefaultendpunct}{\mcitedefaultseppunct}\relax
\EndOfBibitem
\bibitem[Kim \latin{et~al.}(2019)Kim, Lim, Lee, Lee, Kim, Park, Jeon, Park,
  Park, and Cheong]{s41467-018-08284-6}
Kim,~K.; Lim,~S.~Y.; Lee,~J.-U.; Lee,~S.; Kim,~T.~Y.; Park,~K.; Jeon,~G.~S.;
  Park,~C.-H.; Park,~J.-G.; Cheong,~H. Suppression of magnetic ordering in
  XXZ-type antiferromagnetic monolayer NiPS3. \emph{Nat. Commun.}
  \textbf{2019}, \emph{10}, 345\relax
\mciteBstWouldAddEndPuncttrue
\mciteSetBstMidEndSepPunct{\mcitedefaultmidpunct}
{\mcitedefaultendpunct}{\mcitedefaultseppunct}\relax
\EndOfBibitem
\bibitem[Huang \latin{et~al.}(2018)Huang, Clark, Klein, MacNeill,
  Navarro-Moratalla, Seyler, Wilson, McGuire, Cobden, Xiao, Yao,
  Jarillo-Herrero, and Xu]{s41565-018-0121-3}
Huang,~B.; Clark,~G.; Klein,~D.~R.; MacNeill,~D.; Navarro-Moratalla,~E.;
  Seyler,~K.~L.; Wilson,~N.; McGuire,~M.~A.; Cobden,~D.~H.; Xiao,~D.; Yao,~W.;
  Jarillo-Herrero,~P.; Xu,~X. Electrical control of 2D magnetism in bilayer
  CrI3. \emph{Nat. Nano.} \textbf{2018}, \emph{13}, 544--548\relax
\mciteBstWouldAddEndPuncttrue
\mciteSetBstMidEndSepPunct{\mcitedefaultmidpunct}
{\mcitedefaultendpunct}{\mcitedefaultseppunct}\relax
\EndOfBibitem
\bibitem[Jiang \latin{et~al.}(2018)Jiang, Li, Wang, Mak, and
  Shan]{s41565-018-0135-x}
Jiang,~S.; Li,~L.; Wang,~Z.; Mak,~K.~F.; Shan,~J. Controlling magnetism in 2D
  CrI3 by electrostatic doping. \emph{Nat. Nano.} \textbf{2018}, \emph{13},
  549--553\relax
\mciteBstWouldAddEndPuncttrue
\mciteSetBstMidEndSepPunct{\mcitedefaultmidpunct}
{\mcitedefaultendpunct}{\mcitedefaultseppunct}\relax
\EndOfBibitem
\bibitem[Lado and Fern{\'{a}}ndez-Rossier(2017)Lado, and
  Fern{\'{a}}ndez-Rossier]{Lado_2017}
Lado,~J.~L.; Fern{\'{a}}ndez-Rossier,~J. On the origin of magnetic anisotropy
  in two dimensional {CrI}3. \emph{2D Mater.} \textbf{2017}, \emph{4},
  035002\relax
\mciteBstWouldAddEndPuncttrue
\mciteSetBstMidEndSepPunct{\mcitedefaultmidpunct}
{\mcitedefaultendpunct}{\mcitedefaultseppunct}\relax
\EndOfBibitem
\bibitem[Khomskii(2014)]{Khomskii}
Khomskii,~D.~I. \emph{Transition Metal Compounds}; Cambridge University Press,
  2014\relax
\mciteBstWouldAddEndPuncttrue
\mciteSetBstMidEndSepPunct{\mcitedefaultmidpunct}
{\mcitedefaultendpunct}{\mcitedefaultseppunct}\relax
\EndOfBibitem
\bibitem[Cox(2010)]{Cox}
Cox,~P. \emph{Transition Metal Oxides, An introduction to their electronic
  structure and properties}; Mayer J. et al. (Springer), 2010\relax
\mciteBstWouldAddEndPuncttrue
\mciteSetBstMidEndSepPunct{\mcitedefaultmidpunct}
{\mcitedefaultendpunct}{\mcitedefaultseppunct}\relax
\EndOfBibitem
\bibitem[Maeno \latin{et~al.}(2012)Maeno, Kittaka, Nomura, Yonezawa, and
  Ishida]{JPSJ.81.011009}
Maeno,~Y.; Kittaka,~S.; Nomura,~T.; Yonezawa,~S.; Ishida,~K. Evaluation of
  Spin-Triplet Superconductivity in Sr2RuO4. \emph{J. Phys. Soc. Jap.}
  \textbf{2012}, \emph{81}, 011009\relax
\mciteBstWouldAddEndPuncttrue
\mciteSetBstMidEndSepPunct{\mcitedefaultmidpunct}
{\mcitedefaultendpunct}{\mcitedefaultseppunct}\relax
\EndOfBibitem
\bibitem[Jain \latin{et~al.}(2017)Jain, Krautloher, Porras, Ryu, Chen,
  Abernathy, Park, Ivanov, Chaloupka, Khaliullin, Keimer, and Kim]{nphys4077}
Jain,~A.; Krautloher,~M.; Porras,~J.; Ryu,~G.~H.; Chen,~D.~P.;
  Abernathy,~D.~L.; Park,~J.~T.; Ivanov,~A.; Chaloupka,~J.; Khaliullin,~G.;
  Keimer,~B.; Kim,~B.~J. Higgs mode and its decay in a two-dimensional
  antiferromagnet. \emph{Nat. Phys.} \textbf{2017}, \emph{13}, 633\relax
\mciteBstWouldAddEndPuncttrue
\mciteSetBstMidEndSepPunct{\mcitedefaultmidpunct}
{\mcitedefaultendpunct}{\mcitedefaultseppunct}\relax
\EndOfBibitem
\bibitem[Souliou \latin{et~al.}(2017)Souliou, Chaloupka, Khaliullin, Ryu, Jain,
  Kim, Le~Tacon, and Keimer]{PhysRevLett.119.067201}
Souliou,~S.-M.; Chaloupka,~J. c.~v.; Khaliullin,~G.; Ryu,~G.; Jain,~A.;
  Kim,~B.~J.; Le~Tacon,~M.; Keimer,~B. Raman Scattering from Higgs Mode
  Oscillations in the Two-Dimensional Antiferromagnet
  ${\mathrm{Ca}}_{2}{\mathrm{RuO}}_{4}$. \emph{Phys. Rev. Lett.} \textbf{2017},
  \emph{119}, 067201\relax
\mciteBstWouldAddEndPuncttrue
\mciteSetBstMidEndSepPunct{\mcitedefaultmidpunct}
{\mcitedefaultendpunct}{\mcitedefaultseppunct}\relax
\EndOfBibitem
\bibitem[Avdeev \latin{et~al.}(2011)Avdeev, Thorogood, Carter, Kennedy, Ting,
  Singh, and Wallwork]{10.1021/ja109431t}
Avdeev,~M.; Thorogood,~G.~J.; Carter,~M.~L.; Kennedy,~B.~J.; Ting,~J.;
  Singh,~D.~J.; Wallwork,~K.~S. Antiferromagnetism in a Technetium Oxide.
  Structure of CaTcO3. \emph{J. Am. Chem. Soc.} \textbf{2011}, \emph{133},
  1654--1657\relax
\mciteBstWouldAddEndPuncttrue
\mciteSetBstMidEndSepPunct{\mcitedefaultmidpunct}
{\mcitedefaultendpunct}{\mcitedefaultseppunct}\relax
\EndOfBibitem
\bibitem[Rodriguez \latin{et~al.}(2011)Rodriguez, Poineau, Llobet, Kennedy,
  Avdeev, Thorogood, Carter, Seshadri, Singh, and
  Cheetham]{PhysRevLett.106.067201}
Rodriguez,~E.~E.; Poineau,~F.; Llobet,~A.; Kennedy,~B.~J.; Avdeev,~M.;
  Thorogood,~G.~J.; Carter,~M.~L.; Seshadri,~R.; Singh,~D.~J.; Cheetham,~A.~K.
  High Temperature Magnetic Ordering in the $4d$ Perovskite
  ${\mathrm{SrTcO}}_{3}$. \emph{Phys. Rev. Lett.} \textbf{2011}, \emph{106},
  067201\relax
\mciteBstWouldAddEndPuncttrue
\mciteSetBstMidEndSepPunct{\mcitedefaultmidpunct}
{\mcitedefaultendpunct}{\mcitedefaultseppunct}\relax
\EndOfBibitem
\bibitem[Shi \latin{et~al.}(2009)Shi, Guo, Yu, Arai, Belik, Sato, Yamaura,
  Takayama-Muromachi, Tian, Yang, Li, Varga, Mitchell, and
  Okamoto]{PhysRevB.80.161104}
Shi,~Y.~G.; Guo,~Y.~F.; Yu,~S.; Arai,~M.; Belik,~A.~A.; Sato,~A.; Yamaura,~K.;
  Takayama-Muromachi,~E.; Tian,~H.~F.; Yang,~H.~X.; Li,~J.~Q.; Varga,~T.;
  Mitchell,~J.~F.; Okamoto,~S. Continuous metal-insulator transition of the
  antiferromagnetic perovskite ${\rm{NaOsO}}_{3}$. \emph{Phys. Rev. B}
  \textbf{2009}, \emph{80}, 161104\relax
\mciteBstWouldAddEndPuncttrue
\mciteSetBstMidEndSepPunct{\mcitedefaultmidpunct}
{\mcitedefaultendpunct}{\mcitedefaultseppunct}\relax
\EndOfBibitem
\bibitem[Calder \latin{et~al.}(2012)Calder, Garlea, McMorrow, Lumsden, Stone,
  Lang, Kim, Schlueter, Shi, Yamaura, Sun, Tsujimoto, and
  Christianson]{PhysRevLett.108.257209}
Calder,~S.; Garlea,~V.~O.; McMorrow,~D.~F.; Lumsden,~M.~D.; Stone,~M.~B.;
  Lang,~J.~C.; Kim,~J.-W.; Schlueter,~J.~A.; Shi,~Y.~G.; Yamaura,~K.;
  Sun,~Y.~S.; Tsujimoto,~Y.; Christianson,~A.~D. Magnetically Driven
  Metal-Insulator Transition in ${\rm{NaOsO}}_{3}$. \emph{Phys. Rev. Lett.}
  \textbf{2012}, \emph{108}, 257209\relax
\mciteBstWouldAddEndPuncttrue
\mciteSetBstMidEndSepPunct{\mcitedefaultmidpunct}
{\mcitedefaultendpunct}{\mcitedefaultseppunct}\relax
\EndOfBibitem
\bibitem[Wakabayashi \latin{et~al.}(2019)Wakabayashi, Krockenberger, Tsujimoto,
  Boykin, Tsuneyuki, Taniyasu, and Yamamoto]{s41467-019-08440-6}
Wakabayashi,~Y.~K.; Krockenberger,~Y.; Tsujimoto,~N.; Boykin,~T.;
  Tsuneyuki,~S.; Taniyasu,~Y.; Yamamoto,~H. Ferromagnetism above 1000 K in a
  highly cation-ordered double-perovskite insulator Sr3OsO6. \emph{Nat.
  Commun.} \textbf{2019}, \emph{10}, 535\relax
\mciteBstWouldAddEndPuncttrue
\mciteSetBstMidEndSepPunct{\mcitedefaultmidpunct}
{\mcitedefaultendpunct}{\mcitedefaultseppunct}\relax
\EndOfBibitem
\bibitem[Tian \latin{et~al.}(2015)Tian, Svoboda, Ochi, Matsuda, Cao, Cheng,
  Sales, Mandrus, Arita, Trivedi, and Yan]{Nan1}
Tian,~W.; Svoboda,~C.; Ochi,~M.; Matsuda,~M.; Cao,~H.~B.; Cheng,~J.-G.;
  Sales,~B.~C.; Mandrus,~D.~G.; Arita,~R.; Trivedi,~N.; Yan,~J.-Q. High
  antiferromagnetic transition temperature of the honeycomb compound
  ${\mathrm{SrRu}}_{2}{\mathrm{O}}_{6}$. \emph{Phys. Rev. B} \textbf{2015},
  \emph{92}, 100404\relax
\mciteBstWouldAddEndPuncttrue
\mciteSetBstMidEndSepPunct{\mcitedefaultmidpunct}
{\mcitedefaultendpunct}{\mcitedefaultseppunct}\relax
\EndOfBibitem
\bibitem[Hiley \latin{et~al.}(2015)Hiley, Scanlon, Sokol, Woodley, Ganose,
  Sangiao, De~Teresa, Manuel, Khalyavin, Walker, Lees, and Walton]{Hiley2}
Hiley,~C.~I.; Scanlon,~D.~O.; Sokol,~A.~A.; Woodley,~S.~M.; Ganose,~A.~M.;
  Sangiao,~S.; De~Teresa,~J.~M.; Manuel,~P.; Khalyavin,~D.~D.; Walker,~M.;
  Lees,~M.~R.; Walton,~R.~I. Antiferromagnetism at $T> 500$ K in the layered
  hexagonal ruthenate $\mathrm{SrR}{\mathrm{u}}_{2}{\mathrm{O}}_{6}$.
  \emph{Phys. Rev. B} \textbf{2015}, \emph{92}, 104413\relax
\mciteBstWouldAddEndPuncttrue
\mciteSetBstMidEndSepPunct{\mcitedefaultmidpunct}
{\mcitedefaultendpunct}{\mcitedefaultseppunct}\relax
\EndOfBibitem
\bibitem[Okamoto \latin{et~al.}(2017)Okamoto, Ochi, Arita, Yan, and
  Trivedi]{Nan3}
Okamoto,~S.; Ochi,~M.; Arita,~R.; Yan,~J.; Trivedi,~N. Localized-itinerant
  dichotomy and unconventional magnetism in SrRu2O6. \emph{Scientific Reports}
  \textbf{2017}, \emph{7}, 11742\relax
\mciteBstWouldAddEndPuncttrue
\mciteSetBstMidEndSepPunct{\mcitedefaultmidpunct}
{\mcitedefaultendpunct}{\mcitedefaultseppunct}\relax
\EndOfBibitem
\bibitem[Pchelkina \latin{et~al.}(2016)Pchelkina, Streltsov, and Mazin]{mazin}
Pchelkina,~Z.~V.; Streltsov,~S.~V.; Mazin,~I.~I. Spectroscopic signatures of
  molecular orbitals in transition metal oxides with a honeycomb lattice.
  \emph{Phys. Rev. B} \textbf{2016}, \emph{94}, 205148\relax
\mciteBstWouldAddEndPuncttrue
\mciteSetBstMidEndSepPunct{\mcitedefaultmidpunct}
{\mcitedefaultendpunct}{\mcitedefaultseppunct}\relax
\EndOfBibitem
\bibitem[Ochi \latin{et~al.}(2016)Ochi, Arita, Trivedi, and Okamoto]{Nan2}
Ochi,~M.; Arita,~R.; Trivedi,~N.; Okamoto,~S. Strain-induced topological
  transition in ${\mathrm{SrRu}}_{2}{\mathrm{O}}_{6}$ and
  ${\mathrm{CaOs}}_{2}{\mathrm{O}}_{6}$. \emph{Phys. Rev. B} \textbf{2016},
  \emph{93}, 195149\relax
\mciteBstWouldAddEndPuncttrue
\mciteSetBstMidEndSepPunct{\mcitedefaultmidpunct}
{\mcitedefaultendpunct}{\mcitedefaultseppunct}\relax
\EndOfBibitem
\bibitem[Hiley \latin{et~al.}(2014)Hiley, Lees, Fisher, Thompsett, Agrestini,
  Smith, and Walton]{Hiley1}
Hiley,~C.~I.; Lees,~M.~R.; Fisher,~J.~M.; Thompsett,~D.; Agrestini,~S.;
  Smith,~R.~I.; Walton,~R.~I. Ruthenium(V) Oxides from Low-Temperature
  Hydrothermal Synthesis. \emph{Angew. Chem. Int. Ed.} \textbf{2014},
  \emph{53}, 4423\relax
\mciteBstWouldAddEndPuncttrue
\mciteSetBstMidEndSepPunct{\mcitedefaultmidpunct}
{\mcitedefaultendpunct}{\mcitedefaultseppunct}\relax
\EndOfBibitem
\bibitem[Kane and Mele(2005)Kane, and Mele]{Kane}
Kane,~C.~L.; Mele,~E.~J. ${Z}_{2}$ Topological Order and the Quantum Spin Hall
  Effect. \emph{Phys. Rev. Lett.} \textbf{2005}, \emph{95}, 146802\relax
\mciteBstWouldAddEndPuncttrue
\mciteSetBstMidEndSepPunct{\mcitedefaultmidpunct}
{\mcitedefaultendpunct}{\mcitedefaultseppunct}\relax
\EndOfBibitem
\bibitem[Haldane(1988)]{Haldane}
Haldane,~F. D.~M. Model for a Quantum Hall Effect without Landau Levels:
  Condensed-Matter Realization of the "Parity Anomaly". \emph{Phys. Rev. Lett.}
  \textbf{1988}, \emph{61}, 2015--2018\relax
\mciteBstWouldAddEndPuncttrue
\mciteSetBstMidEndSepPunct{\mcitedefaultmidpunct}
{\mcitedefaultendpunct}{\mcitedefaultseppunct}\relax
\EndOfBibitem
\bibitem[Zhang \latin{et~al.}(2005)Zhang, Tan, Stormer, and Kim]{Zhang1}
Zhang,~Y.; Tan,~Y.-W.; Stormer,~H.~L.; Kim,~P. Experimental observation of the
  quantum Hall effect and Berry's phase in graphene. \emph{Nature}
  \textbf{2005}, \emph{438}, 201--204\relax
\mciteBstWouldAddEndPuncttrue
\mciteSetBstMidEndSepPunct{\mcitedefaultmidpunct}
{\mcitedefaultendpunct}{\mcitedefaultseppunct}\relax
\EndOfBibitem
\bibitem[Banerjee \latin{et~al.}(2016)Banerjee, Bridges, Yan, Aczel, Li, Stone,
  Granroth, Lumsden, Yiu, Knolle, Bhattacharjee, Kovrizhin, Moessner, Tennant,
  Mandrus, and Nagler]{Kitaev4}
Banerjee,~A. \latin{et~al.}  Proximate Kitaev quantum spin liquid behaviour in
  a honeycomb magnet. \emph{Nature Materials} \textbf{2016}, \emph{15}, 733 EP
  --, Article\relax
\mciteBstWouldAddEndPuncttrue
\mciteSetBstMidEndSepPunct{\mcitedefaultmidpunct}
{\mcitedefaultendpunct}{\mcitedefaultseppunct}\relax
\EndOfBibitem
\bibitem[Sears \latin{et~al.}(2015)Sears, Songvilay, Plumb, Clancy, Qiu, Zhao,
  Parshall, and Kim]{PhysRevB.91.144420}
Sears,~J.~A.; Songvilay,~M.; Plumb,~K.~W.; Clancy,~J.~P.; Qiu,~Y.; Zhao,~Y.;
  Parshall,~D.; Kim,~Y.-J. Magnetic order in $\alpha-{\mathrm{RuCl}}_{3}$: A
  honeycomb-lattice quantum magnet with strong spin-orbit coupling. \emph{Phys.
  Rev. B} \textbf{2015}, \emph{91}, 144420\relax
\mciteBstWouldAddEndPuncttrue
\mciteSetBstMidEndSepPunct{\mcitedefaultmidpunct}
{\mcitedefaultendpunct}{\mcitedefaultseppunct}\relax
\EndOfBibitem
\bibitem[Choi \latin{et~al.}(2012)Choi, Coldea, Kolmogorov, Lancaster, Mazin,
  Blundell, Radaelli, Singh, Gegenwart, Choi, Cheong, Baker, Stock, and
  Taylor]{Yogesh3}
Choi,~S.~K.; Coldea,~R.; Kolmogorov,~A.~N.; Lancaster,~T.; Mazin,~I.~I.;
  Blundell,~S.~J.; Radaelli,~P.~G.; Singh,~Y.; Gegenwart,~P.; Choi,~K.~R.;
  Cheong,~S.-W.; Baker,~P.~J.; Stock,~C.; Taylor,~J. Spin Waves and Revised
  Crystal Structure of Honeycomb Iridate ${\mathrm{Na}}_{2}{\mathrm{IrO}}_{3}$.
  \emph{Phys. Rev. Lett.} \textbf{2012}, \emph{108}, 127204\relax
\mciteBstWouldAddEndPuncttrue
\mciteSetBstMidEndSepPunct{\mcitedefaultmidpunct}
{\mcitedefaultendpunct}{\mcitedefaultseppunct}\relax
\EndOfBibitem
\bibitem[Jackeli and Khaliullin(2009)Jackeli, and Khaliullin]{Khaliullin}
Jackeli,~G.; Khaliullin,~G. Mott Insulators in the Strong Spin-Orbit Coupling
  Limit: From Heisenberg to a Quantum Compass and Kitaev Models. \emph{Phys.
  Rev. Lett.} \textbf{2009}, \emph{102}, 017205\relax
\mciteBstWouldAddEndPuncttrue
\mciteSetBstMidEndSepPunct{\mcitedefaultmidpunct}
{\mcitedefaultendpunct}{\mcitedefaultseppunct}\relax
\EndOfBibitem
\bibitem[Singh and Gegenwart(2010)Singh, and Gegenwart]{Yogesh1}
Singh,~Y.; Gegenwart,~P. Antiferromagnetic Mott insulating state in single
  crystals of the honeycomb lattice material ${\rm{Na}}_{2}{\rm{IrO}}_{3}$.
  \emph{Phys. Rev. B} \textbf{2010}, \emph{82}, 064412\relax
\mciteBstWouldAddEndPuncttrue
\mciteSetBstMidEndSepPunct{\mcitedefaultmidpunct}
{\mcitedefaultendpunct}{\mcitedefaultseppunct}\relax
\EndOfBibitem
\bibitem[Gupta \latin{et~al.}(2016)Gupta, Sriluckshmy, Mehlawat, Balodhi,
  Mishra, Hassan, Ramakrishnan, Muthu, Singh, and Sood]{Yogesh2}
Gupta,~S.~N.; Sriluckshmy,~P.~V.; Mehlawat,~K.; Balodhi,~A.; Mishra,~D.~K.;
  Hassan,~S.~R.; Ramakrishnan,~T.~V.; Muthu,~D. V.~S.; Singh,~Y.; Sood,~A.~K.
  Raman signatures of strong Kitaev exchange correlations in (Na1-{xLix})2IrO3:
  Experiments and theory. \emph{{EPL} (Europhysics Letters)} \textbf{2016},
  \emph{114}, 47004\relax
\mciteBstWouldAddEndPuncttrue
\mciteSetBstMidEndSepPunct{\mcitedefaultmidpunct}
{\mcitedefaultendpunct}{\mcitedefaultseppunct}\relax
\EndOfBibitem
\bibitem[Knolle \latin{et~al.}(2014)Knolle, Chern, Kovrizhin, Moessner, and
  Perkins]{Knolle}
Knolle,~J.; Chern,~G.-W.; Kovrizhin,~D.~L.; Moessner,~R.; Perkins,~N.~B. Raman
  Scattering Signatures of Kitaev Spin Liquids in ${A}_{2}{\mathrm{IrO}}_{3}$
  Iridates with $A=\mathrm{Na}$ or Li. \emph{Phys. Rev. Lett.} \textbf{2014},
  \emph{113}, 187201\relax
\mciteBstWouldAddEndPuncttrue
\mciteSetBstMidEndSepPunct{\mcitedefaultmidpunct}
{\mcitedefaultendpunct}{\mcitedefaultseppunct}\relax
\EndOfBibitem
\bibitem[Streltsov \latin{et~al.}(2015)Streltsov, Mazin, and
  Foyevtsova]{Streltsov}
Streltsov,~S.; Mazin,~I.~I.; Foyevtsova,~K. Localized itinerant electrons and
  unique magnetic properties of ${\mathrm{SrRu}}_{2}{\mathrm{O}}_{6}$.
  \emph{Phys. Rev. B} \textbf{2015}, \emph{92}, 134408\relax
\mciteBstWouldAddEndPuncttrue
\mciteSetBstMidEndSepPunct{\mcitedefaultmidpunct}
{\mcitedefaultendpunct}{\mcitedefaultseppunct}\relax
\EndOfBibitem
\bibitem[J. and Gemming~T(2008)J., and Gemming~T]{Gemming}
J.,~T.; Gemming~T,~e. b. M. J. e.~a. \emph{Proceedings of the EMC2008: 14th
  European Microscopy Congress}; ISBN 978-3-540-85301-5, 2008\relax
\mciteBstWouldAddEndPuncttrue
\mciteSetBstMidEndSepPunct{\mcitedefaultmidpunct}
{\mcitedefaultendpunct}{\mcitedefaultseppunct}\relax
\EndOfBibitem
\bibitem[Nicolosi \latin{et~al.}(2013)Nicolosi, Chhowalla, Kanatzidis, Strano,
  and Coleman]{Nicolosi1}
Nicolosi,~V.; Chhowalla,~M.; Kanatzidis,~M.~G.; Strano,~M.~S.; Coleman,~J.~N.
  Liquid Exfoliation of Layered Materials. \emph{Science} \textbf{2013},
  \emph{340}\relax
\mciteBstWouldAddEndPuncttrue
\mciteSetBstMidEndSepPunct{\mcitedefaultmidpunct}
{\mcitedefaultendpunct}{\mcitedefaultseppunct}\relax
\EndOfBibitem
\bibitem[Griffin \latin{et~al.}(2018)Griffin, Harvey, Cunningham, Scullion,
  Tian, Shih, Gruening, Donegan, Santos, Backes, and Coleman]{Griffin}
Griffin,~A.; Harvey,~A.; Cunningham,~B.; Scullion,~D.; Tian,~T.; Shih,~C.-J.;
  Gruening,~M.; Donegan,~J.~F.; Santos,~E. J.~G.; Backes,~C.; Coleman,~J.~N.
  Spectroscopic Size and Thickness Metrics for Liquid-Exfoliated h-BN.
  \emph{Chemistry of Materials} \textbf{2018}, \emph{30}, 1998--2005\relax
\mciteBstWouldAddEndPuncttrue
\mciteSetBstMidEndSepPunct{\mcitedefaultmidpunct}
{\mcitedefaultendpunct}{\mcitedefaultseppunct}\relax
\EndOfBibitem
\bibitem[Ogilvie \latin{et~al.}(2017)Ogilvie, Large, Fratta, Meloni,
  Canton-Vitoria, Tagmatarchis, Massuyeau, Ewels, King, and Dalton]{Sean}
Ogilvie,~S.~P.; Large,~M.~J.; Fratta,~G.; Meloni,~M.; Canton-Vitoria,~R.;
  Tagmatarchis,~N.; Massuyeau,~F.; Ewels,~C.~P.; King,~A. A.~K.; Dalton,~A.~B.
  Considerations for spectroscopy of liquid-exfoliated 2D materials: emerging
  photoluminescence of N-methyl-2-pyrrolidone. \emph{Scientific Reports}
  \textbf{2017}, \emph{7}, 16706\relax
\mciteBstWouldAddEndPuncttrue
\mciteSetBstMidEndSepPunct{\mcitedefaultmidpunct}
{\mcitedefaultendpunct}{\mcitedefaultseppunct}\relax
\EndOfBibitem
\bibitem[Backes \latin{et~al.}(2014)Backes, Smith, McEvoy, Berner, McCloskey,
  Nerl, O'Neill, King, Higgins, Hanlon, Scheuschner, Maultzsch, Houben,
  Duesberg, Donegan, Nicolosi, and Coleman]{Backes}
Backes,~C. \latin{et~al.}  Edge and confinement effects allow in situ
  measurement of size and thickness of liquid-exfoliated nanosheets.
  \emph{Nature Communications} \textbf{2014}, \emph{5}, 4576 EP --,
  Article\relax
\mciteBstWouldAddEndPuncttrue
\mciteSetBstMidEndSepPunct{\mcitedefaultmidpunct}
{\mcitedefaultendpunct}{\mcitedefaultseppunct}\relax
\EndOfBibitem
\bibitem[Puthirath~Balan \latin{et~al.}(2018)Puthirath~Balan, Radhakrishnan,
  Woellner, Sinha, Deng, Reyes, Rao, Paulose, Neupane, Apte, Kochat, Vajtai,
  Harutyunyan, Chu, Costin, Galvao, Mart{\'i}, van Aken, Varghese, Tiwary,
  Malie Madom Ramaswamy~Iyer, and Ajayan]{Balan}
Puthirath~Balan,~A. \latin{et~al.}  Exfoliation of a non-van der Waals material
  from iron ore hematite. \emph{Nature Nanotechnology} \textbf{2018},
  \emph{13}, 602--609\relax
\mciteBstWouldAddEndPuncttrue
\mciteSetBstMidEndSepPunct{\mcitedefaultmidpunct}
{\mcitedefaultendpunct}{\mcitedefaultseppunct}\relax
\EndOfBibitem
\bibitem[Osada and Sasaki(2012)Osada, and Sasaki]{Sasaki}
Osada,~M.; Sasaki,~T. Two-Dimensional Dielectric Nanosheets: Novel
  Nanoelectronics From Nanocrystal Building Blocks. \emph{Advanced Materials}
  \textbf{2012}, \emph{24}, 210--228\relax
\mciteBstWouldAddEndPuncttrue
\mciteSetBstMidEndSepPunct{\mcitedefaultmidpunct}
{\mcitedefaultendpunct}{\mcitedefaultseppunct}\relax
\EndOfBibitem
\bibitem[Sun \latin{et~al.}(2014)Sun, Liao, Dou, Hwang, Park, Jiang, Kim, and
  Dou]{Sun}
Sun,~Z.; Liao,~T.; Dou,~Y.; Hwang,~S.~M.; Park,~M.-S.; Jiang,~L.; Kim,~J.~H.;
  Dou,~S.~X. Generalized self-assembly of scalable two-dimensional transition
  metal oxide nanosheets. \emph{Nature Communications} \textbf{2014}, \emph{5},
  3813 EP --, Article\relax
\mciteBstWouldAddEndPuncttrue
\mciteSetBstMidEndSepPunct{\mcitedefaultmidpunct}
{\mcitedefaultendpunct}{\mcitedefaultseppunct}\relax
\EndOfBibitem
\bibitem[Aksit \latin{et~al.}(2012)Aksit, Toledo, and Robinson]{Robinson}
Aksit,~M.; Toledo,~D.~P.; Robinson,~R.~D. Scalable nanomanufacturing of
  millimetre-length 2D NaxCoO2 nanosheets. \emph{J. Mater. Chem.}
  \textbf{2012}, \emph{22}, 5936--5944\relax
\mciteBstWouldAddEndPuncttrue
\mciteSetBstMidEndSepPunct{\mcitedefaultmidpunct}
{\mcitedefaultendpunct}{\mcitedefaultseppunct}\relax
\EndOfBibitem
\bibitem[Ponosov \latin{et~al.}(2019)Ponosov, Komleva, Zamyatin, Walton, and
  Streltsov]{Ponosov}
Ponosov,~Y.~S.; Komleva,~E.~V.; Zamyatin,~D.~A.; Walton,~R.~I.;
  Streltsov,~S.~V. Raman spectroscopy of the low-dimensional antiferromagnet
  ${\mathrm{SrRu}}_{2}{\mathrm{O}}_{6}$ with large N\'eel temperature.
  \emph{Phys. Rev. B} \textbf{2019}, \emph{99}, 085103\relax
\mciteBstWouldAddEndPuncttrue
\mciteSetBstMidEndSepPunct{\mcitedefaultmidpunct}
{\mcitedefaultendpunct}{\mcitedefaultseppunct}\relax
\EndOfBibitem
\bibitem[Pulickel~Ajayan and Banerjee(2016)Pulickel~Ajayan, and Banerjee]{Kim3}
Pulickel~Ajayan,~P.~K.; Banerjee,~K. Two-dimensional van der Waals materials.
  \emph{Physics Today} \textbf{2016}, \emph{69}, 38\relax
\mciteBstWouldAddEndPuncttrue
\mciteSetBstMidEndSepPunct{\mcitedefaultmidpunct}
{\mcitedefaultendpunct}{\mcitedefaultseppunct}\relax
\EndOfBibitem
\bibitem[Kim \latin{et~al.}(2017)Kim, DaSilva, Huang, Fallahazad, Larentis,
  Taniguchi, Watanabe, LeRoy, MacDonald, and Tutuc]{Kim4}
Kim,~K.; DaSilva,~A.; Huang,~S.; Fallahazad,~B.; Larentis,~S.; Taniguchi,~T.;
  Watanabe,~K.; LeRoy,~B.~J.; MacDonald,~A.~H.; Tutuc,~E. Tunable moir{\'e}
  bands and strong correlations in small-twist-angle bilayer graphene.
  \emph{Proceedings of the National Academy of Sciences} \textbf{2017},
  \emph{114}, 3364--3369\relax
\mciteBstWouldAddEndPuncttrue
\mciteSetBstMidEndSepPunct{\mcitedefaultmidpunct}
{\mcitedefaultendpunct}{\mcitedefaultseppunct}\relax
\EndOfBibitem
\bibitem[Yoo \latin{et~al.}(2019)Yoo, Engelke, Carr, Fang, Zhang, Cazeaux,
  Sung, Hovden, Tsen, Taniguchi, Watanabe, Yi, Kim, Luskin, Tadmor, Kaxiras,
  and Kim]{Kim5}
Yoo,~H. \latin{et~al.}  Atomic and electronic reconstruction at the van der
  Waals interface in twisted bilayer graphene. \emph{Nature Materials}
  \textbf{2019}, \emph{18}, 448--453\relax
\mciteBstWouldAddEndPuncttrue
\mciteSetBstMidEndSepPunct{\mcitedefaultmidpunct}
{\mcitedefaultendpunct}{\mcitedefaultseppunct}\relax
\EndOfBibitem
\bibitem[Warner \latin{et~al.}(2009)Warner, Rümmeli, Gemming, Büchner, and
  Briggs]{Warner}
Warner,~J.~H.; Rümmeli,~M.~H.; Gemming,~T.; Büchner,~B.; Briggs,~G. A.~D.
  Direct Imaging of Rotational Stacking Faults in Few Layer Graphene.
  \emph{Nano Lett.} \textbf{2009}, \emph{9}, 102--106\relax
\mciteBstWouldAddEndPuncttrue
\mciteSetBstMidEndSepPunct{\mcitedefaultmidpunct}
{\mcitedefaultendpunct}{\mcitedefaultseppunct}\relax
\EndOfBibitem
\bibitem[Kresse and Hafner(1993)Kresse, and Hafner]{Kresse1}
Kresse,~G.; Hafner,~J. Ab initio molecular dynamics for liquid metals.
  \emph{Phys. Rev. B} \textbf{1993}, \emph{47}, 558--561\relax
\mciteBstWouldAddEndPuncttrue
\mciteSetBstMidEndSepPunct{\mcitedefaultmidpunct}
{\mcitedefaultendpunct}{\mcitedefaultseppunct}\relax
\EndOfBibitem
\bibitem[Kresse and Furthm\"uller(1996)Kresse, and Furthm\"uller]{Kresse2}
Kresse,~G.; Furthm\"uller,~J. Efficient iterative schemes for ab initio
  total-energy calculations using a plane-wave basis set. \emph{Phys. Rev. B}
  \textbf{1996}, \emph{54}, 11169--11186\relax
\mciteBstWouldAddEndPuncttrue
\mciteSetBstMidEndSepPunct{\mcitedefaultmidpunct}
{\mcitedefaultendpunct}{\mcitedefaultseppunct}\relax
\EndOfBibitem
\bibitem[Bl\"ochl(1994)]{Blochl}
Bl\"ochl,~P.~E. Projector augmented-wave method. \emph{Phys. Rev. B}
  \textbf{1994}, \emph{50}, 17953--17979\relax
\mciteBstWouldAddEndPuncttrue
\mciteSetBstMidEndSepPunct{\mcitedefaultmidpunct}
{\mcitedefaultendpunct}{\mcitedefaultseppunct}\relax
\EndOfBibitem
\bibitem[Perdew \latin{et~al.}(1996)Perdew, Burke, and Ernzerhof]{Perdew}
Perdew,~J.~P.; Burke,~K.; Ernzerhof,~M. Generalized Gradient Approximation Made
  Simple. \emph{Phys. Rev. Lett.} \textbf{1996}, \emph{77}, 3865--3868\relax
\mciteBstWouldAddEndPuncttrue
\mciteSetBstMidEndSepPunct{\mcitedefaultmidpunct}
{\mcitedefaultendpunct}{\mcitedefaultseppunct}\relax
\EndOfBibitem
\bibitem[Tian \latin{et~al.}(2015)Tian, Svoboda, Ochi, Matsuda, Cao, Cheng,
  Sales, Mandrus, Arita, Trivedi, and Yan]{Tian}
Tian,~W.; Svoboda,~C.; Ochi,~M.; Matsuda,~M.; Cao,~H.~B.; Cheng,~J.-G.;
  Sales,~B.~C.; Mandrus,~D.~G.; Arita,~R.; Trivedi,~N.; Yan,~J.-Q. High
  antiferromagnetic transition temperature of the honeycomb compound
  ${\mathrm{SrRu}}_{2}{\mathrm{O}}_{6}$. \emph{Phys. Rev. B} \textbf{2015},
  \emph{92}, 100404\relax
\mciteBstWouldAddEndPuncttrue
\mciteSetBstMidEndSepPunct{\mcitedefaultmidpunct}
{\mcitedefaultendpunct}{\mcitedefaultseppunct}\relax
\EndOfBibitem
\bibitem[Suzuki \latin{et~al.}(2019)Suzuki, Gretarsson, Ishikawa, Ueda, Yang,
  Liu, Kim, Kukusta, Yaresko, Minola, Sears, Francoual, Wille, Nuss, Takagi,
  Kim, Khaliullin, Yavas, and Keimer]{Suzuki1}
Suzuki,~H. \latin{et~al.}  Spin waves and spin-state transitions in a ruthenate
  high-temperature antiferromagnet. \emph{Nature Materials} \textbf{2019},
  \relax
\mciteBstWouldAddEndPunctfalse
\mciteSetBstMidEndSepPunct{\mcitedefaultmidpunct}
{}{\mcitedefaultseppunct}\relax
\EndOfBibitem
\bibitem[Singh(2015)]{David}
Singh,~D.~J. Electronic structure and the origin of the high ordering
  temperature in ${\mathrm{SrRu}}_{2}{\mathrm{O}}_{6}$. \emph{Phys. Rev. B}
  \textbf{2015}, \emph{91}, 214420\relax
\mciteBstWouldAddEndPuncttrue
\mciteSetBstMidEndSepPunct{\mcitedefaultmidpunct}
{\mcitedefaultendpunct}{\mcitedefaultseppunct}\relax
\EndOfBibitem
\bibitem[Hariki \latin{et~al.}(2017)Hariki, Hausoel, Sangiovanni, and
  Kune\ifmmode~\check{s}\else \v{s}\fi{}]{Hariki}
Hariki,~A.; Hausoel,~A.; Sangiovanni,~G.; Kune\ifmmode~\check{s}\else
  \v{s}\fi{},~J. DFT+DMFT study on soft moment magnetism and covalent bonding
  in ${\mathrm{SrRu}}_{2}{\mathrm{O}}_{6}$. \emph{Phys. Rev. B} \textbf{2017},
  \emph{96}, 155135\relax
\mciteBstWouldAddEndPuncttrue
\mciteSetBstMidEndSepPunct{\mcitedefaultmidpunct}
{\mcitedefaultendpunct}{\mcitedefaultseppunct}\relax
\EndOfBibitem
\end{mcitethebibliography}

\end{document}